\documentclass[aps,pra,twocolumn,showpacs,preprintnumbers,amsmath,amssymb,footinbib]{revtex4}
\usepackage{graphicx,epsfig}
\usepackage{bm}
\usepackage{dcolumn}
\usepackage{xcolor}
\usepackage[breaklinks=true,colorlinks,citecolor=blue,linkcolor=blue,urlcolor=blue]{hyperref}

\def\nat#1#2#3{Nature {\bf #1}, #2 (#3)}

\def\sc#1#2#3{Science {\bf #1}, #2 (#3)}
\def\prv#1#2#3{Phys. Rev. {\bf #1}, #2 (#3)}
\def\rmp#1#2#3{Rev. Mod. Phys. {\bf #1}, #2 (#3)}
\def\prl#1#2#3{Phys. Rev. Lett. {\bf #1}, #2 (#3)}
\def\pra#1#2#3{Phys. Rev. A {\bf #1}, #2 (#3)}

\def\epjd#1#2#3{Eur. Phys. J. D {\bf #1}, #2 (#3)}

\def\jpb#1#2#3{J. Phys. B: At. Mol. Opt. Phys. {\bf #1}, #2 (#3)}

\def\qso#1#2#3{Quantum Semiclass. Opt. {\bf #1}, #2 (#3)}

\def\pla#1#2#3{Phys. Lett. A {\bf #1}, #2 (#3)}

\def\ajp#1#2#3{Am. J. Phys. {\bf #1}, #2 (#3)}
\def\ejp#1#2#3{Eur. J. Phys. {\bf #1}, #2 (#3)}

\def\noi{\noindent}
\def\bc{\begin{center}}
\def\ec{\end{center}}
\newcommand{\bea}{\begin{equation}}
\newcommand{\eea}{\end{equation}\noi}
\newcommand{\ber}{\begin{eqnarray}}
\newcommand{\eer}{\end{eqnarray}\noi}
\begin{document}
\title{Particle scattering by harmonically trapped Bose and Fermi gases}
\author{Ankita Bhattacharya$^{1,2}$}
\author{Samir Das$^{1}$}
\author{Shyamal Biswas$^{1}$}\email{sbsp [at] uohyd.ac.in}
\affiliation{$^{1}$School of Physics, University of Hyderabad, C.R. Rao Road, Gachibowli, Hyderabad-500046, India\\
$^{2}$Present Address: Institute of Theoretical Physics, TU Dresden, 01069 Dresden, Germany
}

\date{\today}

\begin{abstract}
We have analytically explored the quantum phenomenon of particle scattering by harmonically trapped Bose and Fermi gases with the short ranged (Fermi-Huang $\delta^3_p$ \cite{Fermi-Huang}) interactions among the incident particle and the scatterers. We have predicted differential scattering cross-sections and their temperature dependence in this regard. Coherent scattering even by a single boson or fermion in the finite geometry gives rise to new tool of determining energy eigenstate of the scatterer. Our predictions on the differential scattering cross-sections, can be tested within the present day experimental setups, specially, for (i) 3-D harmonically trapped interacting Bose-Einstein condensate (BEC), (ii) BECs in a double well, and (iii) BECs in an optical lattice.
\end{abstract}


\maketitle
 

\section{Introduction}
In the existing literature, quantum scattering theory is discussed both for classical scatterers (which are either fixed or having classical motions in space \cite{Griffiths}) and quantum scatterers e.g. quantum scattering by atoms, molecules, nuclei, \textit{etc} \cite{Timmermans}. `Particle'\footnote{By `particle', we mean, wave associated with the particle.} can be scattered coherently from each and every point of the region of space of the quantum scatterer if it is fired onto the region, and carries information about the state of the scatterer after being scattered. There are some theoretical discussions on quantum scattering for unfixed quantum scatterer(s) bounded in a region of space, e.g. diffraction of atoms from a standing-wave Schrodinger field \cite{Bodefeld}, scattering of slowly moving atoms by a 3-D harmonically trapped BEC within Bogoliubov-de Gennes formalism \cite{Idziaszek}, particle scattering by a weakly interacting BEC \cite{Wynveen,Poulsen1,Montina,Haring1,Poulsen,Brand}, transport of atoms across interacting BECs in a 1-D optical lattice \cite{Vicencio}, a nondestructive method to probe a complex quantum system using multi-impurity atoms as quantum probes \cite{Streif}, particle scattering by quantum scatterers in restricted geometries \cite{Ankita}, \textit{etc}. 

In none of the previous works, related to the quantum scatterers, temperature dependence of the scattering amplitude or that of the differential scattering cross-section was studied except that for scatterer(s) in box geometries or in array of boxes \cite{Ankita} and for scatterers in 3-D harmonic trap in the thermodynamic limit \cite{Montina}. Thus, we naturally take up discussion on quantum scattering to introduce quantum scattering with quantized motions of the scatterers in thermal equilibrium in finite geometries of harmonic trap as probe for Fermi-Huang $\delta_p^3$ \cite{Fermi-Huang} interactions (among the `incident' particle and the scatterers), which although are easy to deal with, have huge applications in the field of ultra-cold atoms \cite{Busch,Pitaevskii}. We are specially interested in temperature dependence of differential scattering cross-section for scatterers in the harmonically trapped geometry in this regard, as because, thermodynamic properties of ultra-cold gases in harmonic traps are of growing interest \cite{Dalfovo,Bloch,Giorgini,Pitaevskii}.

If a plane wave ($ e^{ikz}$) associated with a free particle (`particle') of a given momentum ($\textbf{p}=\hbar k\hat{k}$) is scattered by a fixed scatterer (at $\textbf{r}=0$) with an interacting potential ($V_{int}(\textbf{r})$), then a spherical wave ($\frac{ e^{ikr}}{r}$) goes out of the scatter with a scattering amplitude ($f(\theta,\phi)$) to a particular direction ($\theta$ and $\phi$) with respect to the initial direction of incidence ($\hat{k}$). If the scatterer is not fixed, say, the scatterer is a particle in a 1-D simple harmonic oscillator ($-\infty<x_0<\infty$), then, according to the superposition principle, the `particle' would be scattered coherently from all the positions ($\{x_0\}$) with the respective probability density $|\psi_n(x_0)|^2$, where $\psi_n(x_0)$ ($n=0,1,2,....$) is the normalized energy eigenstate of the scatterer. In this situation, spherical waves ($\frac{ e^{ikr'}}{r'}$) go out after scattering form all the source (of scattering) points ($\{x_0\}$). All the outgoing spherical waves ($\{\frac{ e^{ikr'}}{r'}\}$) interfere, at a distance $\textbf{r}=\textbf{x}_0+\textbf{r}'$ to a particular direction ($\theta,\phi$) from the center of the oscillation, with different phases and give rise to a coherent scattering amplitude $f_n(\theta,\phi)$ which now depends on the quantum state ($|\psi_n>$) of the scatterer. Small angle neutron scattering by quantum dots was investigated by Pinero \textit{et al} without precisely probing quantized motions of the scatterers in them \cite{Pinero}. Particle scattering by coherent media was also studied experimentally by Chikkatur \textit{et al} \cite{Chikkatur} and Bromley \textit{et al} \cite{Bromley}. While Bromley \textit{et al} did not probe quantized motions of the scatterers in the dense medium, Chikkatur \textit{et al}, could probe quantized motion of the scatterers, to a certain extent, in a BEC; though they did not probe angular dependence of the scattering amplitude. However, electron scattering by harmonically trapped BEC \cite{Wang} and Fermi gas \cite{Wang2} was studied theoretically for $T\rightarrow0$. Although temperature dependence in particle scattering by a BEC was studied by Montina \cite{Montina}, he considered quantized motions of the bosonic scatterers in the thermodynamic limit within Thomas-Fermi approximation. About light scattering by a BEC or by ultracold atoms in optical trap, the experimental work of Schneble \textit{et al} in Ref. \cite{Schneble}, the recent theoretical works of Ezhova \textit{et al} in Ref. \cite{Ezhova}, Zhu \textit{et al} in Ref. \cite{Zhu} and Kozlowski \textit{et al} in Ref. \cite{Kozlowski}, the review work of Mekhov and Ritsch in Ref. \cite{Mekhov}, and the references therein, are quite interesting. Above all, temperature dependence of the differential scattering cross-section, for particle scattering with quantized motion(s) of the scatterer(s) in harmonic/optical trap, has not been studied so far.

This article begins with revisiting of the quantum theory of particle scattering for a fixed (classical) scatterer with Fermi-Huang potential (i.e. regularized $\delta^3$ potential: $V_{int}(\textbf{r})=g\delta_p^3(\textbf{r})=g\delta^3(\textbf{r})\frac{\partial}{\partial r}r$). Then we have generalized the theory for quantum scatterer(s) in restricted geometries, in particular, for bosonic/fermionic scatterer(s) in a (i) 1-D harmonic trap, (ii) 2-D harmonic trap, and (iii) 3-D harmonic trap. Then we have calculated the scattering amplitudes, and have plotted the differential scattering cross-sections for all the cases. We also have investigated temperature dependence of the differential scattering cross-sections for the above cases, and specially emphasized on the differential cross-section for particle scattering by BEC(s) in the 3-D harmonic trap \cite{Davis}, double well trap, and the optical lattice.

\section{Particle scattering by a single scatterer in a harmonic trap}
In quantum scattering theory of particle scattering we deal with the time independent Schrodinger equation
\begin{eqnarray}\label{eqn:1}
\bigg(-\frac{\hbar^2}{2m}\nabla^2+V_{int}(\textbf{r})\bigg)\psi(\textbf{r})=E\psi(\textbf{r}).
\end{eqnarray}
Scattering of an incident particle of mass $m$ and the given momentum $\hbar k\hat{k}$, is recast, as scattering of a `particle' (i.e. scattering of the plane wave $\psi_{in}\equiv e^{ikz}$) by the interacting potential $V_{int}(\textbf{r})$, into an outgoing spherical wave $\psi_{out}\equiv\frac{ e^{ikr}}{r}$. General form of the solution to Eqn.(\ref{eqn:1}), in the radiation zone, takes the form \cite{Griffiths}
\begin{eqnarray}\label{eqn:2}
\psi(\textbf{r})=\psi(r,\theta,\phi)\simeq A\bigg[{ e^{ikz}+f(\theta,\phi)\frac{ e^{ikr}}{r}}\bigg],
\end{eqnarray}
where $|A|^2$ is proportional to the intensity of the incident `particle'. From this information, we can find out the scattering (probability) amplitude ($f(\theta,\phi)$) of the out going spherical wave to a particular direction ($\theta,\phi$ in usual convention) with respect to the direction of the incidence. The scattering amplitude, for $V_{int}(\textbf{r})=g\delta_p^3(\textbf{r})$, with all orders of the Born series, takes the form \cite{Notes,Cavalcanti,BDR,Pitaevskii}
\begin{eqnarray}\label{eqn:5}
f(\theta,\phi)=-\frac{mg}{2\pi\hbar^2(1+ik\frac{mg}{2\pi\hbar^2})}.
\end{eqnarray}
We have considered the scatterer to be fixed for Eqn.(\ref{eqn:5}). If the scatterer is not fixed, rather having a relative motion with the incident `particle' keeping the interacting potential unaltered, then the scattering amplitude would take the form
\begin{eqnarray}\label{eqn:6}
f(\theta,\phi)=-\frac{\bar{\mu}g}{2\pi\hbar^2(1+ik\frac{\bar{\mu}g}{2\pi\hbar^2})},
\end{eqnarray}
where $\bar{\mu}=\frac{mM}{m+M}$ is the reduced mass and $M$ is the mass of the scatterer \cite{Pitaevskii}. The scattering amplitude is independent of $\theta$ and $\phi$ for low energy scattering, so that, s-wave scattering length can be conveniently defined, for low energy scattering, as $a_s=\lim_{k\rightarrow0}-f(\theta,\phi)$. Thus, we quantify the coupling constant, as $g=\frac{2\pi\hbar^2a_s}{\bar{\mu}}$. 

\subsection{For a single scatterer in a 1-D harmonic trap}
If $\textbf{r}$ be the position of the incident particle, such that the center of the trapped potential ($\textbf{r}=\textbf{0}$) is the origin, then the $\delta_p^3$ interaction between the incident particle at $\textbf{r}$ and the scatterer at $x_0\hat{i}$ can be expressed as  
\begin{eqnarray}\label{eqn:8}
V_{int}(\textbf{r})=g\delta_p^3(\textbf{r}-x_0\hat{i}).
\end{eqnarray}
Eqn.(\ref{eqn:5}) can be recast for this problem by using Eqn.(\ref{eqn:8}) as
\begin{eqnarray}\label{eqn:9}
f(\theta,\phi)=-\frac{mg_k}{2\pi\hbar^2} e^{i(\textbf{k}-\textbf{k}')\cdot x_0\hat{i}}=-\frac{mg_k}{2\pi\hbar^2} e^{-i\textbf{k}'\cdot x_0\hat{i}}
\end{eqnarray}
where $g_k=\frac{g}{1+ika_sm/\bar{\mu}}$. Eqn.(\ref{eqn:9}) is correct only if the scattering has happened only from $\textbf{r}=x_0\hat{i}$, and will not be correct if $x_0\hat{i}$ is not a fixed point. 

\begin{figure}
\includegraphics[width=.98 \linewidth]{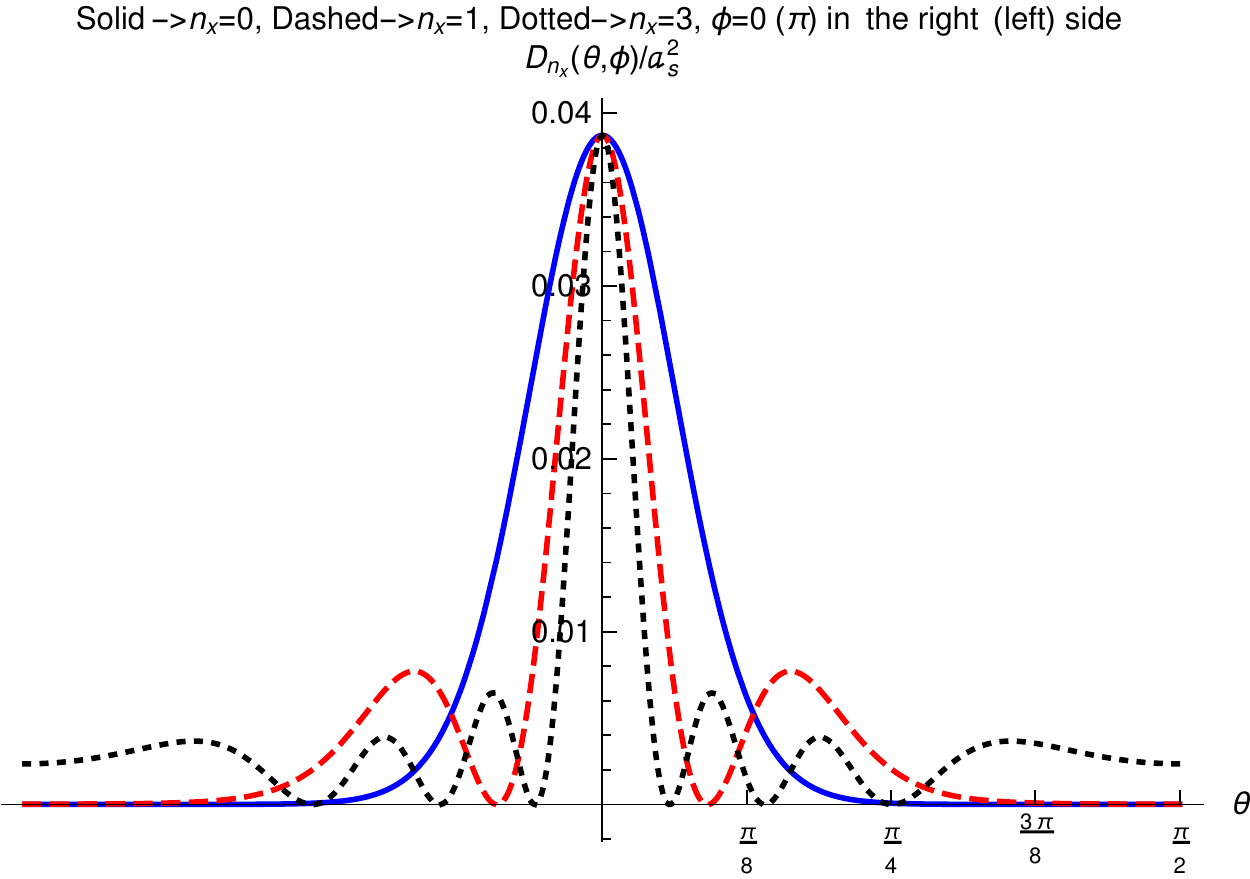}
\caption{Intensity distribution ($D_{n_x}(\theta,\phi)=|f_{n_x}(\theta,\phi)|^2$) along a line parallel to the $x$-axis for scattering of a `particle' ($ e^{ikz}$) by the 1-D harmonic oscillator along $x$-axis for $a_s$ as unit length, $ka_s=5$, $l_x/a_s=1$, and $m/M=0.1$. Plots follow from Eqn.(\ref{eqn:41}).
\label{fig2}}
\end{figure}

Although we can consider Eqn.(\ref{eqn:9}) where there is a relative motion (between the particle and scatterer) by replacing mass of the `particle' by the reduced mass, yet the scatterer is still classical as we have not quantized the motion of the scatterer. Let us now consider quantized motion of the scatterer(s) into the theory of quantum scattering, and begin with the scatterer as a particle in a 1-D harmonic trap potential $V(\textbf{r}_0)=\frac{1}{2}M\omega_x^2x_0^2$ where $\omega_x$ is the angular frequency of oscillations, $M$) is the mass and $\textbf{r}_0=x_0\hat{i}$ is the position of the scatterer such that $-\infty<x_0<\infty$. We again consider the scatterer to scatter the incident `particle', $A e^{ikz}$, by the interacting potential $V_{int}(\textbf{r})=g\delta_p^3(\textbf{r}-x_{0}\hat{i})$. Though the incident `particle' and the scatterer are charge-less, they are distinguished by their mass and spin. The incident `particle' is having zero spin, and it does not feel the trap potential except the contact potential with the scatterer. The scatterer is of nonzero spin (and magnetic moment), so that, it can be trapped by an inhomogeneous magnetic field with the potential energy $V(\textbf{r}_0)=\frac{1}{2}M\omega_x^2x_0^2$ \cite{Pitaevskii}. Scattering amplitude, if the scatterer is fixed at $\textbf{r}=x_0\hat{i}$, would be the same as that in Eqn.(\ref{eqn:9}) as $f(\theta,\phi)=-\frac{mg_k}{2\pi\hbar^2} e^{-i\textbf{k}'\cdot x_0\hat{i}}$. Normalized energy eigenstate of the scatterer, corresponding to the energy eigenvalue $E_{n_x}=(n_x+1/2)\hbar\omega_x$, can be written, as \cite{Griffiths}
\begin{eqnarray}\label{eqn:40}
\psi_{n_x}^{}(x_0)=\sqrt{\frac{1}{\sqrt{\pi}l_x}}\frac{1}{\sqrt{2^{n_x}n_x!}}H_{n_x}(x_0/l_x) e^{-x^2/2l_x^2},
\end{eqnarray}
where $l_x=\sqrt{\hbar/M\omega_x}$ is the confining length scale of the scatterer, and $H_{n_x}(x_0/l_x)$ is the Hermite polynomial of degree $n_x=0,1,2,...$. Now, the quantum scattering is happening from all the points $-\infty<x_0<\infty$ simultaneously with respective probability density $\{|\psi_n(x_0)|^2\}$. Thus, the scattering amplitude for the scatterer in the quantum state $|\psi_n>$, can be written, using Eqn.(\ref{eqn:9}), as
\begin{eqnarray}\label{eqn:41}
f_{n_x}(\theta,\phi)&=&-\frac{mg_k}{2\pi\hbar^2}\int_{-\infty}^{\infty} e^{-i\textbf{k}'\cdot x_0\hat{i}}|\psi_{n_x}(x_0)|^2dx_0\nonumber\\&=&-\frac{mg_k}{2\pi\hbar^2} e^{-q_x^2l_x^2}L_{n_x}(2q_x^2l_x^2),
\end{eqnarray}
where $q_x=k\sin(\theta)\cos(\phi)/2$ as defined before, and $L_{n_x}(2q_x^2l_x^2)$ is the Laguerre polynomial of degree $n_x$ \cite{Gradshteyn}. We show the profile of the differential scattering cross-section ($D_{n_x}(\theta,\phi)=|f_{n_x}(\theta,\phi)|^2$) for the 1-D case in FIG. \ref{fig2} for different quantum numbers.

\subsection{For a single scatterer in a 2-D harmonic trap}
For 2-D case, the trap potential would be $V(\textbf{r}_0)=\frac{1}{2}M\omega_x^2x_0^2+\frac{1}{2}M\omega_y^2y_0^2$ where $\omega_y$ is the angular frequency of oscillations along $y$ direction, and $\textbf{r}_0=x_0\hat{i}+y_0\hat{j}$ is the position of the scatterer such that $-\infty<y_0<\infty$. We again consider, that, the scatterer to scatter the incident `particle', $A e^{ikz}$, by the interacting potential $V_{int}(\textbf{r})=g\delta_p^3(\textbf{r}-x_{0}\hat{i}-y_{0}\hat{j})$. Thus, scattering amplitude, for the 2-D case, would be, in the separable form
\begin{eqnarray}\label{eqn:42}
f^{}_{n_x,n_y}(\theta,\phi)=-\frac{mg_k}{2\pi\hbar^2} e^{-q_x^2l_x^2-q_y^2l_y^2}L_{n_x}(2q_x^2l_x^2)L_{n_y}(2q_y^2l_y^2)
\end{eqnarray}
where $\psi_{n_x,n_y}^{}(x_0,y_0)$ is the normalized energy eigenstate of the scatterer with energy eigenvalue $E_{n_x,n_y}=(n_x+1/2)\hbar\omega_x+(n_y+1/2)\hbar\omega_y$ and $n_y=0,1,2,...$.

\subsection{For a single scatterer in a 3-D harmonic trap}
Above generalization, however, is not obvious for the scatterer in a 3-D harmonic trap potential $V(\textbf{r}_0)=\frac{1}{2}M\omega_x^2x_0^2+\frac{1}{2}M\omega_y^2y_0^2+\frac{1}{2}M\omega_z^2z_0^2$ as because we further have to consider momentum transfer mechanism for the motion of the scatterer along the $z$ direction since the incident `particle' has momentum only along the $z$ direction. For this reason, generalization Eqn.(\ref{eqn:9}), for an arbitrary fixed position $\textbf{r}_0=x_0\hat{i}+y_0\hat{j}+z_0\hat{k}$ in 3-D, would be $f(\theta,\phi)=-\frac{mg_k}{2\pi\hbar^2} e^{-i\textbf{k}'\cdot(x_0\hat{i}+y_0\hat{j})+i(\textbf{k}-\textbf{k}')\cdot z_0\hat{k}}$. Thus, 3-D generalization of Eqn.(\ref{eqn:42}) would be in the separable form
\begin{eqnarray}\label{eqn:43}
f_{n_x,n_y,n_z}(\theta,\phi)&=&-\frac{mg_k}{2\pi\hbar^2}\int e^{-i\textbf{k}'\cdot(x_0\hat{i}+y_0\hat{j})+i(\textbf{k}-\textbf{k}')\cdot z_0\hat{k}}\nonumber\\&&\times|\psi_{n_x,n_y,n_z}^{}(x_0,y_0,z_0)|^2d^3\textbf{r}_0\nonumber\\&=&-\frac{mg_k}{2\pi\hbar^2} e^{-q_x^2l_x^2-q_y^2l_y^2-\bar{q}_z^2l_z^2}\nonumber\\&&\times L_{n_x}(2q_x^2l_x^2)L_{n_y}(2q_y^2l_y^2)L_{n_z}(2\bar{q}_z^2l_z^2),~~~~~
\end{eqnarray}
where $\psi_{n_x,n_y,n_z}^{}(x_0,y_0,z_0)$ is the normalized energy eigenstate of the quantum scatterer in the 3-D harmonic trap with energy eigenvalue $E_{n_x,n_y,n_z}=(n_x+1/2)\hbar\omega_x+(n_y+1/2)\hbar\omega_y+(n_z+1/2)\hbar\omega_z$, $n_z=0,1,2,...$, $\omega_z$ is the angular frequency of oscillation of the scatterer along the $z$ direction, $l_z=\sqrt{\hbar/M\omega_z}$, and $\bar{q}_z=-k(1-\cos\theta)/2=-k\sin^2(\theta/2)$ which acts like an obliquity factor. Differential scattering cross-section for the 3-D harmonic scatterer, can be obtained from Eqn.(\ref{eqn:43}), as
\begin{eqnarray}\label{eqn:44}
D_{n_x,n_y,n_z}(\theta,\phi)&=&\bigg|\frac{mg_k}{2\pi\hbar^2} e^{-q_x^2l_x^2-q_y^2l_y^2-\bar{q}_z^2l_z^2}\nonumber\\&&\times L_{n_x}(2q_x^2l_x^2)L_{n_y}(2q_y^2l_y^2)L_{n_z}(2\bar{q}_z^2l_z^2)\bigg|^2.~~~~~
\end{eqnarray}
Eqn.(\ref{eqn:44}) though goes beyond the first Born approximation, it is fully consistent (for $ka_s\ll1$)
with the result obtained by Bodefeld and Wilkens after truncating the Lippmann-Schwinger equation to the level of the first Born approximation \cite{Bodefeld}. Averaging over the position of the scatterer in Eqn. (\ref{eqn:43}) (and that in the preceding two as well) is justified by the fundamental principle of superposition\footnote{The superposition principle is often applied in a similar way for the light scattering (diffraction) by a double slit. Please see R. P. Feynman, R. B. Leighton, and M. L. Sands, \textit{The Feynman Lectures on Physics: Quantum Mechanics}, Vol. 3, Chapter 1, Addison-Wesley, MA (1965) for the same.}, that, if we do not know the initial position of the scatterer rather know only its energy eigenstate $|\psi_{n_x,n_y,n_z}>$, then the scattering takes place from all the points $\{\textbf{r}_0\}$ of the scatterer with the respective probability densities $\{|\psi_{n_x,n_y,n_z} (x_0,y_0,z_0 )|^2\}$. We are considering the energy eigenstate $|\psi_{n_x,n_y,n_z}>$ to be unaltered in the process of scattering. Energy eigenstate would change in the process of inelastic scattering \cite{Bodefeld}. We will discuss about the reasons in the concluding section to justify less probability of the inelastic scattering in the context of thermal and many-body effects \cite{Idziaszek}.

\section{Particle scattering by Bose and Fermi gases in thermodynamic equilibrium in 3-D harmonic traps}
Let us now consider $N$ identical ideal scatterers in the 3-D harmonic trap \cite{Dalfovo,Pitaevskii}. Above expression in Eqn.(\ref{eqn:43}) can be generalized for these scatterers, all of which scatter the incident `particle' ($A e^{ikz}$) by the same delta potential ($V_{int}(\textbf{r})=\sum_{j=1}^{N}g\delta_p^3(\textbf{r}-\textbf{r}_{0j})$), as
\begin{eqnarray}\label{eqn:45}
f_{\textbf{n}_1,\textbf{n}_2,...,\textbf{n}_N}(\theta,\phi)&=&-\frac{mg_k}{2\pi\hbar^2} e^{-||\bar{\textbf{q}}\cdot\textbf{l}||^2}\sum_{\textbf{n}_{j=1}}^{\textbf{n}_{j=N}}L_{n_{jx}}(2q_x^2l_x^2)\nonumber\\&&\times L_{n_{jy}}(2q_y^2l_y^2)L_{n_{jz}}(2\bar{q}_z^2l_z^2),~~~~~~
\end{eqnarray}
where $\textbf{n}_j=(n_{jx},n_{jy},n_{jz})$ represents quantum numbers corresponding the energy eigenstate of the $j$th oscillator, and $||\bar{\textbf{q}}\cdot\textbf{l}||^2=q_x^2l_x^2+q_y^2l_y^2+\bar{q}_z^2l_z^2$. Eqn.(\ref{eqn:45}), however, is applicable not only for distinguishable scatterers, but also for Bose and Fermi scatterers as all the energy eigenstates are orthogonal.

\begin{figure}
\includegraphics[width=0.98 \linewidth]{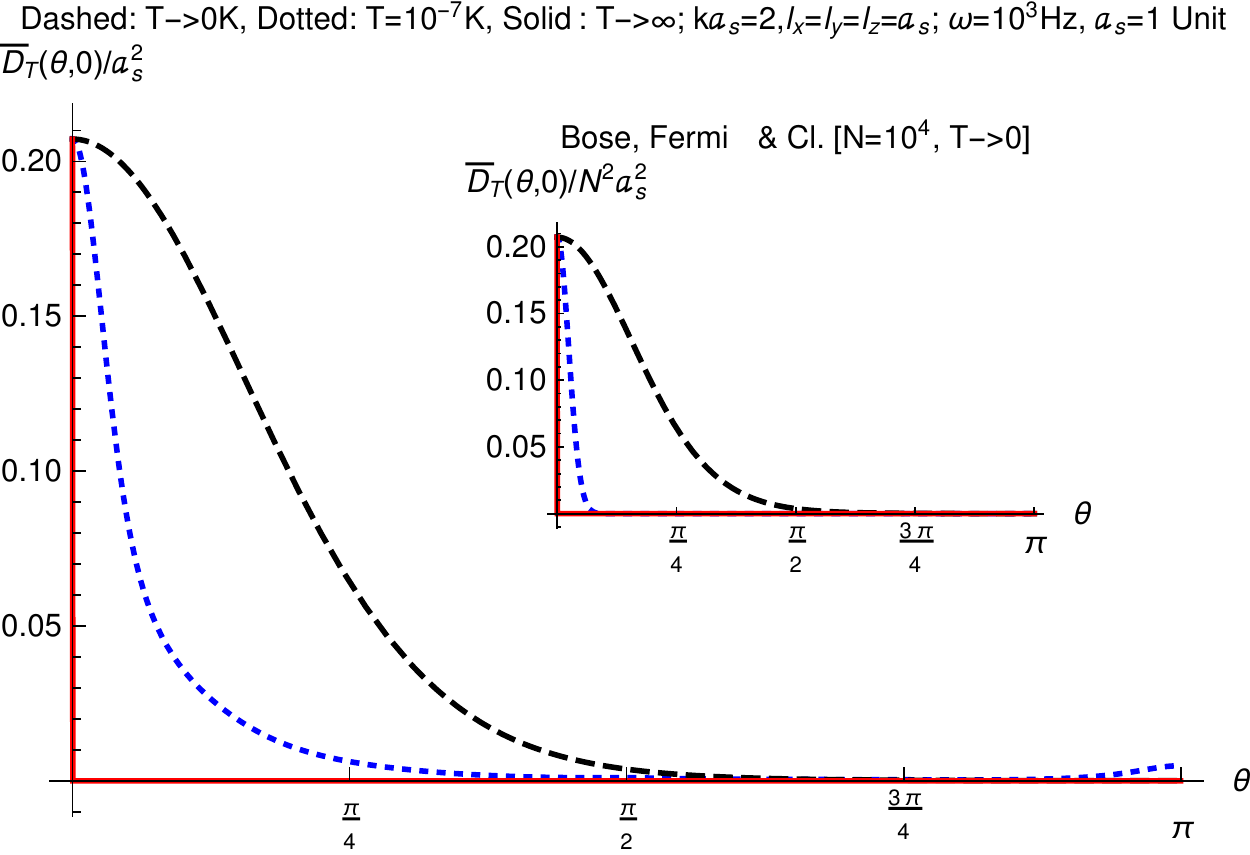}
\caption{Intensity distribution for scattering of a `particle' ($ e^{ikz}$) by a 3-D isotropic harmonic oscillator for $a_s$ as unit length and $m/M=0.1$. Plots follow from the right hand side of Eqn.(\ref{eqn:46}) with $\bar{n}_\textbf{n}$ replaced by $P_n$. Dashed, dotted and solid lines correspond to $D_{T}(\theta,0)$ for $T\rightarrow0$, $T\rightarrow10^{-7}$K, and $T\rightarrow\infty$, respectively. In the inset, the parameters remain same, except the temperature and no. scatterers. Dashed, dotted and solid lines in the inset are linked to Bose gas (Eqn.(\ref{eqn:47})), Fermi gas (Eqn.(\ref{eqn:47a})), and classical scatterers (Eqn.(\ref{eqn:46}) for $\textbf{n}\rightarrow\infty$ limit) in the 3-D isotropic harmonic trap. 
\label{fig3}}
\end{figure}

Let us now consider the ideal scatterers in thermodynamic equilibrium with its surroundings at temperature $T$ and chemical potential $\mu$. Scattering amplitude for the scatterers would now depend upon the temperature and chemical potential, and can be written, as
\begin{eqnarray}\label{eqn:46}
\bar{f}_{T}(\theta,\phi)&=&-\frac{mg_k}{2\pi\hbar^2} e^{-||\bar{\textbf{q}}\cdot\textbf{l}||^2}\sum_{\textbf{n}=(0,0,0)}^{(\infty,\infty,\infty)}\bar{n}_{\textbf{n}}L_{n_{x}}(2q_x^2l_x^2)\nonumber\\&&\times L_{n_{y}}(2q_y^2l_y^2)L_{n_{z}}(2\bar{q}_z^2l_z^2),
\end{eqnarray}
where $\bar{n}_\textbf{n}=\frac{1}{ e^{(E_{\textbf{n}}-\mu)/k_BT}\mp1}$ represents no. of scatterers in the single-particle quantum state $\psi_\textbf{n}(\textbf{r}_0)=\psi_{n_x,n_y,n_z}(x_0,y_0,z_0)$ for Bose ($-$) or Fermi ($+$) scatterers, and $E_{\textbf{n}}=E_{n_x,n_y,n_z}=(n_x+1/2)\hbar\omega_x+(n_y+1/2)\hbar\omega_y+(n_z+1/2)\hbar\omega_z$. Eqn.(\ref{eqn:46}) is our prediction for the scattering amplitude for a harmonically trapped ideal Bose or Fermi gas at any temperature. For a single particle, $\bar{n}_\textbf{n}$ in Eqn.(\ref{eqn:46}) can be replaced by the Boltzmann probability $P_{\textbf{n}}= e^{-E_{\textbf{n}}/k_BT}/Z$ where $Z=\sum_{\textbf{n}} e^{-E_{\textbf{n}}/k_BT}$ is the partition function. We show temperature dependence of $\bar{D}_{T}(\theta,\phi)=|\bar{f}_{T}(\theta,\phi)|^2$ for a single particle in FIG. \ref{fig3}. We also show its statistics dependence in the FIG. \ref{fig3} (inset).

For $T\rightarrow0$, all ($N$) the Bose scatterers occupy the ground state. Differential scattering cross-section, in this situation, takes the form, from Eqn.(\ref{eqn:46}), as
\begin{eqnarray}\label{eqn:47}
\bar{D}_{T\rightarrow0}(\theta,\phi)=|\bar{f}_{T\rightarrow0}(\theta,\phi)|^2=|N a_k|^2 e^{-2||\bar{\textbf{q}}\cdot\textbf{l}||^2},
\end{eqnarray}
where $a_k=\frac{mg_k}{2\pi\hbar^2}=\frac{a_sm/\bar{\mu}}{1+ika_sm/\bar{\mu}}$. We plot Eqn.(\ref{eqn:46}) in FIG.~\ref{fig4}(b) for relevant values of parameters. Eqn.(\ref{eqn:47}) leads to the scattering cross-section, for $k\rightarrow0$, as
\begin{eqnarray}\label{eqn:47c-s}
\sigma=\int_0^\pi d\theta\int_0^{2\pi}d\phi\bar{D}_{T\rightarrow0}(\theta,\phi)\sin\theta=4\pi|N a_sm/\bar{\mu}|^2.~~
\end{eqnarray}

On the other hand, for $T\rightarrow0$, all the ($N$) Fermi scatterers (of the same spin component, say spin up) will occupy the first $N$ single particle states. Thus, for large $N$ and isotropic case, modulus squared of the r.h.s. of Eqn.(\ref{eqn:46}), takes the form\footnote{As because, we can approximate $\sum_{n=0}^\infty e^{-n/N}L_n(x)=\frac{ e^{\frac{1}{N}+\frac{x}{1- e^{\frac{1}{N}}}}}{-1+ e^{\frac{1}{N}}}$ for $N\gg1$, as $\sum_{n=0}^NL_n(x)\approx N e^{-Nx}$.}, for harmonically trapped Fermi gas as 
\begin{eqnarray}\label{eqn:47a}
\bar{D}_{T\rightarrow0}(\theta,\phi)=|N a_k|^2 e^{-6||\bar{\textbf{q}}\cdot\textbf{l}||^2N^{1/3}}.
\end{eqnarray}
For classical scatterers ($n_x,n_y,n_z\rightarrow\infty$), in contrary to the above, the scattering amplitude in Eqn.(\ref{eqn:46}) would be infinitely narrow\footnote{As because, for $n\gg1$, we can write $L_n(x)\rightarrow e^{-nx}$ from the expression of the polynomial itself.} as shown in FIG.~\ref{fig3} (inset).

\subsection{Weak interparticle interactions and finite size effects for Bose scatterers in a 3-D harmonic trap}
\begin{figure}
\includegraphics[width=0.98 \linewidth]{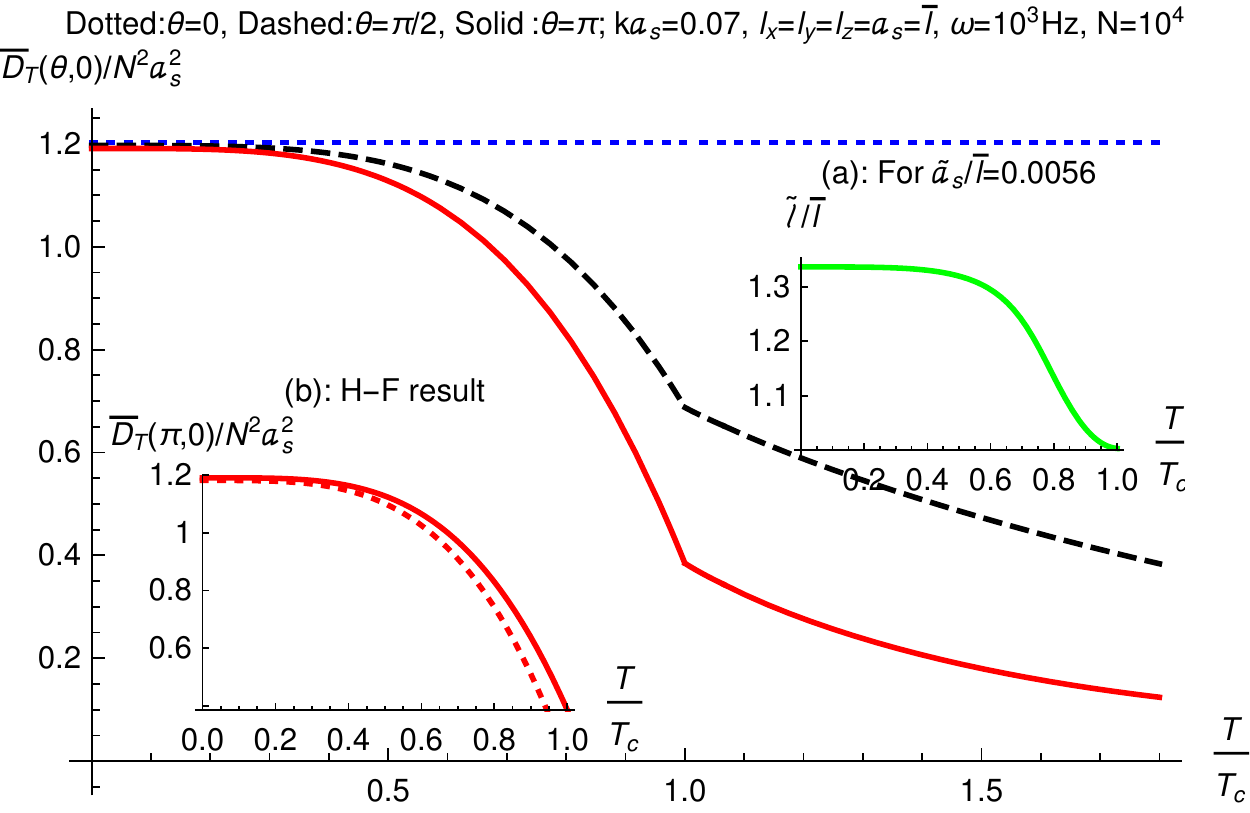}
\caption{Temperature dependence of the differential scattering cross-section along the forward ($\theta=0$, dotted line), perpendicular ($\theta=\pi/2$, dashed line) and backward ($\theta=\pi$, solid lines) directions for 3-D harmonically trapped isotropic ideal Bose gas for the relevant parameters as mentioned above. Plots follow from Eqn.(\ref{eqn:46}) for $m/M=0.1$. Inset-a represents finite temperature scaling ($\bar{l}\rightarrow\tilde{\ell}$) of $\bar{l}$ for the same system (within the 4th order in $\tilde{\ell}/\bar{l}-1$ in the H-F energy functional \cite{Biswas}) for the coupling constant $\frac{4\pi\hbar^2\tilde{a}_s}{M}$ with $\tilde{a}_s=90 a_0=0.0056\bar{l}$ for $^{87}$Rb atoms \cite{Ensher}. Dotted line in the inset-b represents finite size and inter-scatterer effects within the H-F approximation over the solid line which also represents backward scattering in the main figure.  
\label{fig7}}
\end{figure}

Temperature dependence of the scattering amplitude comes from the triple summation in Eqn.(\ref{eqn:46}). The summation, in the thermodynamic limit, followed by the Taylor expansions of the Laguerre polynomials about $k=0$ with $q^2=(q_x^2+q_y^2+\bar{q}_z^2)$, $\omega_x=\omega_y=\omega_z=\omega$, $\bar{l}=\sqrt{\hbar/M\omega}$, $t=\frac{k_BT}{\hbar\omega}$, $z=e^{\mu/k_BT}$ and $N=t^3Li_{3}(z)$ takes the form $S=t^3Li_{3}(z)-6q^2\bar{l}^2t^4Li_{4}(z)+(q^2\bar{l}^2)^2[12t^5Li_{5}(z)+3t^4Li_{4}(z)]+{\it{O}}(q^6)$ for the Bose gas above the condensation point ($T>T_c=\frac{\hbar\omega}{k_B}[N/\zeta(3)]^{1/3}$)\footnote{Here, $Li_{j}(z)=z+\frac{z^2}{2^j}+\frac{z^3}{3^j}+....$ is a poly-logarithmic function of the argument $z$ and order $j$. It is also known as a Bose-Einstein integral.}. For $T<T_c$, similar form also appears with non-condensate fraction $(t/t_c)^3$ where $t_c=\frac{k_BT_c}{\hbar\omega}$. Condensate part shows temperature dependence only in the form of the condensate fraction $\frac{N_0}{N}=1-(t/t_c)^3$. Temperature dependence of $\bar{D}_T(\theta,\phi)=|\bar{f}_T(\theta,\phi)|^2$, for the Bose gas, with the appropriate temperature dependence of the chemical potential \cite{Biswas2}, is shown in FIG. \ref{fig7}. For Fermi gas, only change would be the replacement of the Bose-Einstein integrals ($Li_j(z)$) by the Fermi integrals ($-Li_j(-z)$) $\forall j$. For anisotropic trap, forms of the bulk quantities are mostly unaltered with the replacement $\omega=(\omega_x\omega_y\omega_z)^{1/3}$.

For the finite size of the trap and weak inter-scatterer interactions ($\frac{4\pi\hbar^2\tilde{a}_s}{M}\sum_{i,j<i}\delta_p^3(\vec{r}_{0i}-\vec{r}_{0j})$), the condensate fraction, to the lowest order in $\tilde{a}_s$, takes the form $\frac{N_0}{N}=1-(t/t_c)^3-\frac{3t^2\zeta(2)}{2t_c^3\zeta(3)}-\frac{4.932t^{7/2}\tilde{a}_s}{t_c^3\zeta(3)\bar{l}}$ within the Hartree-Fock (H-F) approximation \cite{Pitaevskii,Giorgini2,Biswas3}. Inter-scatterer interactions do not greatly modify Eqn.(\ref{eqn:46}) as $N^{1/6}\tilde{a}_s/\bar{l}\ll1$. These interactions, apart from modifying the condensate fraction, can substantially scale ($\bar{l}\rightarrow\tilde{\ell}$) the typical confining length $\bar{l}$ (i.e. $l_x$, $l_y$ and $l_z$) in the exponent in Eqn.(\ref{eqn:46}) keeping its form unaltered. Finite temperature scaling of $\bar{l}$, as prescribed in Ref. \cite{Biswas}, is shown in FIG. \ref{fig7} (inset-a) for repulsive interactions. With both the modifications, we have shown corrections due to the finite size and the inter-scatterer interactions effects\footnote{Both the effects are comparable for $0\lnsim T \lnsim T_c$.} to the temperature dependence of $\bar{D}_T(\theta,\phi)$ specially for the backward scattering below $T_c$ in FIG. \ref{fig7} (inset-b). From the trend of the scaling, one can neglect the effect of interactions for $T>T_c$. However, effect of the interactions, for $T\rightarrow0$, may not necessarily be perturbative, and can be better described within Thomas-Fermi approximation \cite{Idziaszek}.

In the FIG.~\ref{fig4}(b) we show the scaling result for the angular dependence ($\theta$) of the differential scattering cross-section for the weakly interacting case of the BEC for $T\rightarrow0$, and finite temperature and size effects over this result within the Hartree-Fock approximation according to the prescription described above. It is quite clear from the plots in the FIG.~\ref{fig4}(b), that repulsive interactions lead to narrowing down the profile of the differential scattering cross-section around $\theta=0$ as the condensate broadens up around $\theta=0$. This is quite natural, as because, the scattering amplitude for the extended object (BEC) is Fourier decomposed at all the source points of scattering. However, if temperature increases, probability of excited states being occupied by the scatterers increases, which in turn increases probability of scattering to some larger angles like that shown in FIG.~\ref{fig2}. Thus, increase of temperature leads to large angle scattering. However, coherency get reduced if scatterers are found in different energy eigen states other than the ground state at a finite temperature. It results reduction of the scattering cross-section with the increase of temperature. This is true in general. This is also apparent in FIG. \ref{fig3} both for ideal Bose and Fermi scatterers in harmonic traps. We will also investigate the same for interacting BECs in other trapped geometries like double-well trap and optical lattice trap.

\section{Particle scattering by Bose scatters in other 3-D optical traps}

\subsection{For Bose scatterers in a double-well potential}
Let us now consider an ideal gas of $N+N$ Bose scatterers in a 3-D double-well potential $V(\textbf{r}_0)=-M\omega_x^2x_0^2/2+M\omega_x^2x_0^4/4d^2+2M\omega_y^2y_0^2/2+2M\omega_z^2z_0^2/2$, such that, frequency of oscillation is the same as that in the previous case, and the minima of double-well are separated along $x$-axis by a distance $d$ \cite{Andrews,Pitaevskii}. In thermodynamic equilibrium, for $T\rightarrow0$, all the particles condense to the ground state.  Within the tight-binding approximation (which is very good for $d\gg l_x$), there would be two distinct condensates of $N$ scatterers in each well, such that each of the condensates scatters the incident `particle' ($A e^{ikz}$) like that in Eqn.(\ref{eqn:47}). However, net scattering amplitude would be the superposition of the scattering amplitudes corresponding to the individual condensate as the setup is analogue of the double slit experiment \cite{Pitaevskii,Bloch2}. Thus, scattering from the two condensates would interfere, as
\begin{eqnarray}\label{eqn:48}
\bar{D}_{T\rightarrow0}(\theta,\phi)=|N a_k|^2 e^{-2||\bar{\textbf{q}}\cdot\textbf{l}||^2}\bigg[2\cos\big(\frac{\pi d\sin(\theta)}{\lambda}\big)\bigg]^2.
\end{eqnarray}
Here we did not consider any Josephson oscillation as $d\gg l_x$ \cite{Levy,Raghavan}. We plot the differential cross-section in FIG.~\ref{fig4}(c) for relevant values of parameters. In the same figure we further present scaling results for weakly interacting Bose scatterers in the double-well trap well below the condensation point and finite temperature and size effects within the Hartree-Fock approximation on top of the tight binding approximation in a similar way as prescribed in the previous section for the Bose scatterers in the harmonic trap.

\subsection{For Bose scatterers in a 1-D optical lattice}
\begin{figure}
\includegraphics[width=0.98 \linewidth]{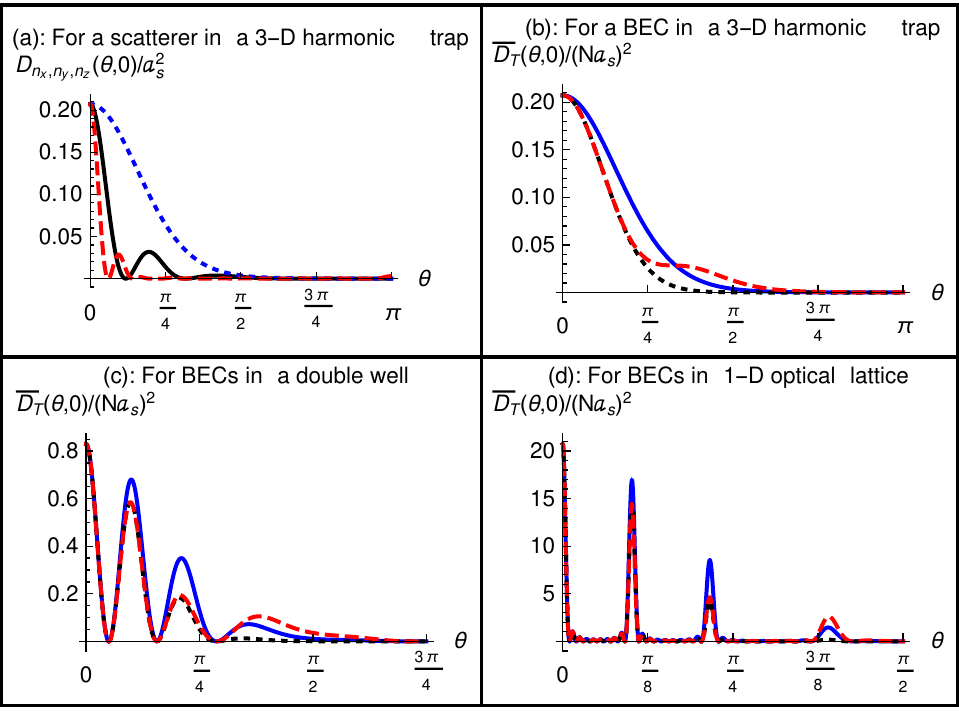}
\caption{Total differential scattering cross-section for quantum scatterer(s) in trapped geometry. For all the figures, we have considered the following: $l_x=l_y=l_z=a_s$, $ks_s=2$, and $m/M=0.1$. Solid, dotted and dashed lines in FIG. \ref{fig4} (a) follow Eqn.(\ref{eqn:44}) for $n_x=5,~n_y=1,~n_z=0$; $n_x=0,~n_y=0,~n_z=0$; and $n_x=20,~n_y=20,~n_z=20$ respectively. Solid line in FIG. \ref{fig4} (b) represents Eqn.(\ref{eqn:47}), the solid line in FIG. \ref{fig4} (c) represents Eqn.(\ref{eqn:48}) for $d=10~l_x$, and the solid line in FIG. \ref{fig4} (d) represents Eqn.(\ref{eqn:49}) for $d=10~l_x$ and $N'=10$. While the solid lines in the last three (\ref{fig4} b,c,d) figures represent non interacting BEC(s), the dotted lines in the same figures represent scaling results for interacting BEC(s) with $\tilde{a}_s=0.0056\bar{l}$ as set in FIG. \ref{fig7}, and dashed lines represent finite temperature and size effects over the dotted lines within the H-F approximation for $T/T_c=0.1$.
\label{fig4}}
\end{figure}

Let us now consider $N'$ 3-D noninteracting BECs in a 1-D optical lattice \cite{Anderson,Bloch}, such that two consecutive condensates are separated along $x$-axis by the lattice spacing $d$. Entire system is in thermodynamic equilibrium. For $T\rightarrow0$, all the condensates have the same ($N$) number of particles. So, the system essentially is a 1-D grating of 3-D condensates. Within the tight-binding approximation, there would be $N'$ distinct condensates of $N$ scatterers in each well, such that each of the condensates scatters the incident `particle' ($A e^{ikz}$) like that in Eqn.(\ref{eqn:47}) \cite{Pedri,Pitaevskii}. However, net scattering amplitude would be the superposition of the scattering amplitudes corresponding to the individual condensate as the setup is now analogue of the 1-D grating experiment. Thus, scattering from the $N'$ condensates would interfere, as
\begin{eqnarray}\label{eqn:49}
\bar{D}_{T\rightarrow0}(\theta,\phi)=|N a_k|^2 e^{-2||\bar{\textbf{q}}\cdot\textbf{l}||^2}\bigg[\frac{\sin(\frac{N'\pi d\sin(\theta)}{\lambda})}{\sin(\frac{\pi d\sin(\theta)}{\lambda})}\bigg]^2.
\end{eqnarray}
Since $d\gg l_x$, Eqn.(\ref{eqn:49}) is good for the Mott insulator phase of the condensates. We plot $\bar{D}_{T\rightarrow0}(\theta,\phi)$ in FIG.~\ref{fig4}(d) for relevant values of parameters. In the same figure we further present scaling results for weakly interacting Bose scatterers in the optical lattice trap well below the condensation point and finite temperature and size effects within the Hartree-Fock approximation on top of the tight binding approximation in a similar way as prescribed in the previous section for the Bose scatterers in the harmonic trap.

Again we see, in FIGs.~\ref{fig4}(c) and (d), according to our expectation, that, repulsive interactions lead to narrowing down the profile of the differential scattering cross-section around $\theta=0$ as the condensates broaden up around $\theta=0$. Increase of temperature, as expected and explained before, leads to large angle scattering also for the scatterers in the double-well trap and the optical lattice trap. Coherency would be lost in presence of the disorders in the BECs. In this situation the differential scattering cross-sections in the FIGs.~\ref{fig4}(b), (c) and (d) would be infinitesimally narrow like that shown by the solid lines in the FIG. \ref{fig3}.

\section{Conclusions}
To conclude, we have presented quantum theory of particle scattering by quantum scatterers in quantized bound states in harmonically trapped geometry for Fermi-Huang $\delta_p^3$ \cite{Fermi-Huang} interactions (between the incident particle and the scatterers), which although are easy to deal with have huge applications in the field of ultra-cold atoms \cite{Busch,Pitaevskii}. Particle scattering by the quantum scatterer(s) in thermal equilibrium in finite geometry of optical traps has not been investigated before us except for $T\rightarrow0$ \cite{Idziaszek,Wang,Wang2}. Temperature dependence of the differential scattering cross-sections, as shown in FIG. \ref{fig3}, would be an important tool to distinguish type (bosonic/fermionic) of the scatterers. The discontinuities in the slops of $\bar{D}_T(\theta,\phi)$s at $T=T_c$, except for the forward scattering as shown in FIG. \ref{fig7}, can be used to detect occurrence of BEC by particle scattering method. Our predictions can be tested within the present day experimental setups.

Just by looking into the scattering intensity-pattern for sufficiently large energy of the incident `particle', as shown in FIG. \ref{fig2}, and counting the maximum number of the zeros of the differential scattering cross-section along $x$-axis, one can easily determine energy eigenstate of 1-D harmonic oscillator, as number of the zeros along the $x$ axis is equal to quantum number $n_x$. From the highest possible peak height of the forward differential scattering cross-section for a scatterer in the harmonic oscillator, one can easily determine scattering length ($a_s$) of the incident `particle' as the height, for low energy of the incident `particle', is proportional to $a_s^2$.      

We have constructed our theory for a single incident `particle'. For a beam of $\bar{N}$ incident `particles', $A$ in Eqn.(\ref{eqn:2}) would be replaced by $\sqrt{\bar{N}}A$, and all the results which depend on `$A$' would be scaled accordingly. However, the scattering amplitude, the differential scattering cross-section, and the total scattering cross-section are independent of `$A$'. So, all our result would be unaltered under this scaling.

Parameters used for plotting the figures are not specific to a particular scattering problem. However, we set $m/M=0.1$ which would be appropriate for $^{40}$K as (fermionic) scatterer and $^{4}$He as the scattered particle. The ratio of $m/M$ though would be even less ($0.046$) for the combination of $^{87}$Rb (bosonic scatterer) and $^{4}$He (scatterer particle) our results would not change much, as $m/\bar{\mu}$ for both the cases are approximately $0.91$ and $0.96$ respectively. We set $a_s/\bar{l}=1$ and $\tilde{a}_s/\bar{l}=0.0056$ (which is appropriate for $^{87}$Rb atoms) to show stronger effect due to the particle scattering than that due to inter-scatterer interactions. Values of $\omega$ and $ka_s$ are set $1000$ and $2$ respectively to clearly show effect of temperature on particle scattering by a harmonic oscillator in the ultra-cold regime ($T\sim10^{-7}$K). If $k$ increases, number of maxima and minima increases in the profile of the differential scattering cross-section. The number of maxima and minima further increases if the quantum number (i.e. the nodes in the wave function of the scatterer) increases. We set $N=10^4$ to show a significant difference between the particle scattering by a Bose gas and that by a Fermi gas in a harmonic trap. The later one shifts towards the classical limit if $N$ increases.

Here, we have considered only elastic scattering. Elementary excitations over the BEC leads to inelastic scattering involving inelastic processes where the trapped particles in scattering out-states are found in different harmonic oscillator states than those in the scattering in-states. Hence, inelastic scattering is less probabilistic at finite temperatures. Moreover, differential scattering cross-section in inelastic channels decays exponentially with the number of scatterers beyond a certain value \cite{Idziaszek}.

Particle scattering by weakly interacting harmonically trapped BEC was already studied, for $T\rightarrow0$, by Idziaszek \textit{et al} with consideration of the first Born approximation for $g\delta^3(\textbf{r})$ potential \cite{Idziaszek}. One may suspect their result, as, $g\delta^3(\textbf{r})$ can not truly scatter a `particle' except in 1-D \cite{Cavalcanti,BDR}. However, the first Born approximation, for $g\delta^3(\textbf{r})$ interaction, surprisingly gives correct result for $ka_s\rightarrow0$. 

Within the last two decades, a lot of experimental observations have been done on harmonically trapped ultracold Bose and Fermi gases. Our prediction of the scattering amplitudes or differential scattering cross-sections in Eqns.(\ref{eqn:46}) to (\ref{eqn:49}) (or that represented in FIGs. \ref{fig3}, \ref{fig7} and \ref{fig4}) may open interests to the experimentalists to study temperature dependence in particle scattering by harmonically trapped Bose and Fermi gases.

Our theory can be generalized, without much difficulty, for weakly interacting scatterer(s) in box geometry with further consideration of scattering (diffraction) by the aperture \cite{Ankita} within perturbative formalism. Our work can be further extended with the consideration of the elementary excitations as prescribed in Ref. \cite{Idziaszek} specially for the condensates in a double well and optical lattice not only for elastic collisions but also for inelastic collisions. However, how to generalize our result for strongly interacting scatterers, e.g. atoms in Feshbach resonance, is an open problem. We consider condensates to be well separated in both the cases of double well and optical lattice. Generalization of results for the Josephson oscillations \cite{Levy,Raghavan} and superfluid phase specially around superfluid-Mott insulator transition \cite{Greiner,Bloch} are kept as open problems. In quantum theory, refraction can be thought of a quantum scattering of a `particle'. In future, our theory can be extended towards the quantum theory of refractive index of a medium of quantum fluid. 

\acknowledgments
S. Das acknowledges financial support (JRF) of the UGC, India. S. Biswas acknowledges financial support of the DST, Govt. of India under the INSPIRE Faculty Award Scheme [No. IFA-13 PH-70].  We are indebted to Prof. C. Timm, TU Dresden, Germany for his valuable critical comments. We are thankful to the reviewers for their thorough reviews and highly appreciate their comments and suggestions, which significantly contributed to improving the quality of the paper. Useful comments from Prof. J.K. Bhattacharjee, IACS, India are also gratefully acknowledged. We also thank Prof. K. Rzazewski, CTPPAS, Poland for introducing the Ref. \cite{Idziaszek} to our knowledge.

\end{document}